\documentclass[a4paper,fleqn,usenatbib]{mnras}
\usepackage{amsmath}
\usepackage{txfonts}

\usepackage[T1]{fontenc}
\usepackage{ae,aecompl}

\usepackage{graphicx}	
\usepackage{amssymb}	

\defcitealias{Bek+17}{Paper~I}
\newcommand{\PaperI}{\citetalias{Bek+17}}

\defcitealias{Bek+18}{Paper~II}
\newcommand{\PaperII}{\citetalias{Bek+18}}

\newcommand{\msun}{\mathcal{M}_\odot}

\newcommand{\tx}[1][]{\mathrm{#1}}

\newcommand{\sfe}{\mathrm{SFE_{gl}}}
\newcommand{\fb}{F_\mathrm{{bound}}}
\newcommand{\rhrt}{\lambda}
\newcommand{\tvr}{t_\mathrm{VR}}

\title[Tidal field impact independent cluster survival]{The star cluster survivability after gas expulsion is independent of the impact of the Galactic tidal field}

\author[B. Shukirgaliyev et al.]{
B. Shukirgaliyev$^{1,2,3}$\thanks{E-mail: \href{mailto: bekdaulet@aphi.kz}{bekdaulet@aphi.kz}}\thanks{Fellow of the International Max-Planck Research School for Astronomy and Cosmic Physics at the University of Heidelberg (IMPRS-HD)}, 
G. Parmentier$^{1}$, 
P. Berczik$^{4,5,1}$ 
and A. Just$^{1}$\\
\\
$^{1}$Zentrum f\"ur Astronomie der Universit\"at Heidelberg, Astronomisches Rechen-Institut, M\"onchhofstr. 12-14, 69120 Heidelberg, Germany\\
$^{2}$Fesenkov Astrophysical Institute, Observatory str. 23, 050020 Almaty, Kazakhstan\\
$^{3}$Faculty of Physics and Technology, Al-Farabi Kazakh National University, Al-Farabi ave. 71, 050040 Almaty, Kazakhstan\\
$^{4}$The International Center of Future Science of the Jilin University,
2699 Qianjin St., 130012, Changchun City, China\\
$^{5}$Main Astronomical Observatory, National Academy of Sciences of Ukraine,
27 Akademika Zabolotnoho St, 03680 Kyiv, Ukraine
}

\date{Accepted XXX. Received YYY; in original form ZZZ}

\pubyear{2018}

\begin{document}
\label{firstpage}
\pagerange{\pageref{firstpage}--\pageref{lastpage}}
\maketitle

\begin{abstract}

We study the impact of the tidal field on the survivability of star clusters following instantaneous gas expulsion. 
Our model clusters are formed with a centrally-peaked {star-formation efficiency profile as a result of star-formation taking place with a constant efficiency per free-fall time.} 
We define the impact of the tidal field as the ratio of the cluster half-mass radius to its Jacobi radius {immediately after gas expulsion}, $\lambda = r_\mathrm{h}/R_\mathrm{J}$.
We vary $\lambda$ by varying {either the Galactocentric distance, or the size (hence volume density) of star clusters.}

{We propose a new method to measure the violent relaxation duration, in which we compare the total mass-loss rate of star clusters with their stellar evolutionary mass-loss rate. That way, we can robustly estimate the bound mass fraction of our model clusters at the end of violent relaxation. The duration of violent relaxation correlates linearly with the Jacobi radius, when considering identical clusters at different Galactocentric distances. In contrast, it is nearly constant for the solar neighbourhood clusters, slightly decreasing with $\lambda$. } The violent relaxation does not last longer than 50 Myr in our simulations.

Identical model clusters placed at different Galactocentric distances {have the same final bound fraction, despite experiencing different impacts} of the tidal field. The solar neighbourhood clusters with different densities {experience only limited variations of their final bound fraction.}

In general, we conclude that the cluster survivability after instantaneous gas expulsion, as measured by their bound mass fraction at the end of violent relaxation, $F_\mathrm{bound}$, is independent of the impact of the tidal field, $\lambda$.  

\end{abstract}




\section{Introduction}

Controversial results have been presented by observers regarding the dependence of the cluster dissolution time on the cluster mass and environment \citep[see][for an overview]{Lamers09,Whitmore17}.
One group of observers reported that star cluster dissolution depends on the cluster mass and cluster environment, i.e. the cluster dissolution time is longer for higher cluster masses and weaker tidal field of the host galaxy \citep{BL03,Lamers+05b,Bastian+12}. However, \citet{Whitmore+07}, \citet{Fall+09}, \citet{Chandar+10}, \citet{Chandar+14} reported from their extragalactic observations that star clusters dissolve independently of their mass and of their environment {within the first billion or half a billion years of their evolution.}

In 2015, the Legacy Extragalactic UV Survey (LEGUS) collaboration started its work to investigate the connection between environmental conditions in galaxies and their cluster populations \citep{Calzetti+15}. One of the aims of the LEGUS collaboration is to investigate the influence of the environment on the cluster evolution/dissolution in nearby galaxies. \citet{Messa+18} study the cluster population of the M51 galaxy as a function of galactocentric distance, and location with respect to the spiral arms (i.e spiral arms and inter-arm regions). Considering the clusters younger than 200 Myr they conclude that the 
cluster age distribution depends on both galactocentric distance and ambient density, showing evidence for faster cluster dissolution in the inner and denser regions than in the outer and diffuse (inter arm) ones, under the assumption of a constant rate of cluster formation. Clusters younger than 10 Myr were not accounted in the age distribution function because their census is contaminated by quickly dissolving unbound clusters. 

{$N$-body simulations of the long-term evolution of initially virialized star clusters show that the dissolution of clusters depends on both the cluster mass and the environment \citep[e.g.][among others]{FH95,BM03,Spurzem+05,TF05,Renaud+08,Rossi+16}. 
The impact of the tidal field is usually measured by the ratio between the cluster half-mass radius and the cluster Jacobi radius (i.e. tidal radius), $\lambda=r_\mathrm{h}/R_\mathrm{J}$. 
\citet{FH95} showed that clusters become unstable for $\lambda \sim 0.7$ and get suddenly disrupted. \citet{Ernst+15}, considering initially virialized solar neighbourhood clusters with different Roche volume filling factors (hence different $\lambda$), showed that Roche volume overfilling clusters dissolve in a mass-independent regime. They argued that star clusters might overfill their Roche lobe as a consequence of residual star-forming gas expulsion.}

{Stellar feedback of high-mass stars can drive the residual gas out of a star-forming region with speeds of about 10 km s$^{-1}$, even for small star-formation efficiencies \citep[SFE$<0.3$,][]{Dib+13,Rahner+19}. When a gas embedded cluster loses most of its mass contained in gas due to gas expulsion, the weakening of its total gravitational potential drives it out of virial equilibrium, thus triggering violent relaxation.}

{By definition, violent relaxation is the dynamical evolution of a star cluster from a state of non-equilibrium into a new state of (quasi-)equilibrium \citep{Lynden-Bell67}. In $N$-body simulations of the long-term evolution of star clusters, the violent relaxation is usually neglected or considered to be over. \citet{Bek+18} (hereafter \PaperII) studied the dissolution of solar neighbourhood clusters starting their $N$-body simulations from the beginning of violent relaxation and showed that the outcome of the latter plays a significant role on the subsequent long-term evolution of clusters.}

{Various aspects of star cluster violent relaxation have been studied: the impact of the gas expulsion time-scale \citep[e.g.][]{GB01,BK07,Brinkmann+17}, the impact of hierarchical star formation \citep[e.g.][]{Smith+13,LeeGoodwin16,Farias+17}, the interplay of gas dynamics with stellar dynamics \citep[e.g.][]{Farias+18,Wall+19}.}

Theoretical works dedicated to the violent relaxation, usually neglect the impact of the tidal field of the host galaxy. Only a few papers in the literature have considered the effect of the tidal field on the cluster early evolution and survivability after gas expulsion in their $N$-body simulations: \citet{Goodwin97,Kroupa+01,BK07,Bek+17}. \citet{BK07} are the only ones who mapped the parameter space of global SFE, gas expulsion time-scale and impact of the tidal field in a comprehensive way. {They reported that the external tidal field has a significant influence on the cluster survivability when $\lambda \gtrsim 0.05$. }

In \citet{Bek+17} (hereafter \PaperI) we studied the survivability of star clusters after instantaneous gas expulsion, assuming that clusters form according to the local-density-driven cluster formation model of \citet{PP13}. The semi-analytical cluster formation model of \citet{PP13} considers a star formation process happening with a constant SFE per free-fall time in a centrally concentrated, spherically symmetric molecular gas clump. As a consequence, the stellar component of the gas embedded cluster has a steeper density profile than the gas component. The results of their cluster formation model also explains the star-formation relation between the surface densities of gas and young stellar objects observed in eight nearby molecular clouds by \citet{Gutermuth+11}. \PaperI\ reports that star clusters which form with such a centrally peaked SFE profile are more resilient to instantaneous gas expulsion than earlier models \citep[e.g.][and references therein]{BK07}. That is, our model clusters survive instantaneous gas expulsion with a critical global SFE of $\sfe=0.15$ instead of $\sfe=0.33$ as estimated previously for monolithically formed star clusters. By global SFE, $\sfe$, we understand the fraction of gas of a star-forming clump converted into stars by the time of instantaneous gas expulsion.

\PaperI\ also investigated the impact of the tidal field by varying the size of clusters on a given orbit and found that it is not significant within the uncertainty\footnote{The uncertainty is a consequence of the randomization of the initial conditions of the $N$-body simulations.} of 10\% on the bound mass fraction retained by star clusters at the end of their violent relaxation. 
However, only a small number of simulations were performed and they were limited to only two different birth masses and 3 realizations per model.
Therefore in this paper we study the impact of the tidal field in a more comprehensive way expanding the results of \PaperI\ to larger ranges of cluster masses and realizations, and also varying the cluster Galactocentric distance $R_\mathrm{orb}$.
This will be helpful for the understanding and interpretation of extragalactic observations.

In section \ref{sec:model} we describe our models, initial conditions and simulations. We measure the duration of violent relaxation in section \ref{sec:VR}. The main results about cluster survivability are presented in section \ref{sec:FB} and the conclusions in section \ref{sec:conc}.

\section{Description of cluster models} \label{sec:model}

\subsection{Cluster initial conditions}

All our model clusters have a Plummer density profile describing their stellar component immediately before instantaneous gas expulsion. We recover the density profile of the residual star-forming gas assuming that star clusters form with a constant SFE per free-fall time ($\epsilon_{\mathrm{ff}}=0.05$) according to the local-density-driven cluster formation model of \citet{PP13}. 
As a consequence, the residual gas volume density profile is shallower than that of the stars (see Fig. 2 in \PaperI). 
We assume that gas embedded clusters are in virial equilibrium with the residual gas potential before gas expulsion.
The initial phase-space distribution of the stars of our model clusters has been generated with the \textsc{mkhalo} program from \texttt{falcON} package of \citet{McMillanDehnen07}, combined to a specially developed external potential plug-in `\textsc{GasPotential}' \citep[\PaperI,][]{BekThesis}. 
{We do not consider the gravitational potential of the residual gas in our $N$-body simulations, assuming that gas expulsion is instantaneous. 
Our model clusters become super-virial once gas expulsion has taken place, because they were in virial equilibrium with the total (gas+stars) gravitational potential.} 
The initial stellar mass function (IMF) of \citet{K2001} with initial stellar mass limits of $m_\mathrm{low}=0.08\msun$ and $m_\mathrm{up}=100\msun$ has been applied. The direct $N$-body simulations are performed with high-resolution paralleled $\phi$-\textsc{grape-gpu} code \citep{Berczik+13} with the \textsc{SSE} \citep{Hurley+00} stellar evolution recipes turned on.

\subsection{The tidal field of the Galaxy}

We consider star clusters on circular orbits in the Galactic disk plane. For the Galactic tidal field we use an axisymmetric three-component Plummer-Kuzmin model \citep{MiyamotoNagai75} with the parameters as given in \citet{Just+09}. For the sake of clarity we provide here the Equation (32) of \citet{Just+09} describing the Galactic tidal field components
\begin{equation}
\Phi(R,z)=-\frac{GM}{\sqrt{R^2 + \left(a + \sqrt{b^2 + z^2} \right)^2}},
\label{eq:GalPot}
\end{equation}
where 
$G$ is the gravitational constant, $M$ is the mass of the component, and $a$ and $b$ represent the flattening and the core radius of the component. Their numerical values are given in Table \ref{tab:GalPot}. The rotation curve obtained from the Galactic potential model is presented in Fig. \ref{fig:vrot}.
\begin{table} 
 \centering
 \caption{The numerical values of the Galaxy component parameters from Eq. \ref{eq:GalPot}.}
 \label{tab:GalPot}
  \begin{tabular}{llrr}
  \hline
  Galaxy component & $M\ [\msun]$ & $a\ [\tx[kpc]]$ &  $b\ [\tx[kpc]]$ \\
  \hline
  Bulge & $1.4\times10^{10}$ & 0.0 & 0.3 \\
  Disk  & $9.0\times10^{10}$ & 3.3 & 0.3 \\
  Halo  & $7.0\times10^{11}$ & 0.0 & 25.0 \\
  \hline
  \end{tabular}
\end{table} 

We also use Equation (13) from \citet{Just+09} to calculate the Jacobi radius:
\begin{equation}
R_\mathrm{J} = \left( \frac{ GM_\mathrm{J} }{ (4-\beta^2)\Omega^2 } \right)^{1/3},
\label{eq:rj}
\end{equation}
where $M_\mathrm{J}$ is the Jacobi mass of the cluster (which is the stellar mass enclosed within one Jacobi radius), $\beta=1.37$ is the normalized epicyclic frequency and $\Omega=V_\mathrm{orb}/R_\mathrm{orb}$ is the angular speed of a star cluster moving  with an orbital speed $V_\mathrm{orb}$ on a circular orbit at a Galactocentric distance $R_\mathrm{orb}$.

{ We consider the Jacobi mass $M_{\tx[J]}$ as the cluster bound mass when we measure the bound fraction of our model clusters. Although it has been discussed that there are stars beyond the Jacobi radius staying around the cluster for several mega-years \citep{Ross+97, Just+09}, we decide to keep our definition of the bound mass as the Jacobi mass to avoid any overestimation.}

\subsection{Parameter space covered by our grid of simulations}

In our previous works (\PaperI\ and \PaperII) we considered the evolution of clusters after instantaneous gas expulsion for different global SFEs ($\sfe=[0.1,0.25]$) and different cluster birth masses $M_\star=[3\tx[k],100\tx[k]]\ \msun$ (i.e. cluster stellar mass at the time of instantaneous gas expulsion). 
We studied both their violent relaxation and their long-term evolution till their final dissolution in the Galactic tidal field. 
All our model clusters have circular orbits in the Galactic disk plane at the Galactocentric distance of $R_\mathrm{orb}=8\ \tx[kpc]$.
{Only model clusters with the impact of the tidal field set to $\lambda=0.052$ were calculated until full dissolution in \PaperII. Hereafter, when we refer to the impact of the tidal field, we use $\lambda=r_\mathrm{h}/R_\mathrm{J}$ measured at $t=0$ (i.e. immediately after instantaneous gas expulsion).}

Now we expand our set of $N$-body simulations and we study the impact of the tidal field on star cluster survivability after instantaneous gas expulsion. We consider our previous models with $\rhrt=0.052$ as our `standard' set of models, or `S0-models'. 

We limit ourselves to clusters that survive instantaneous gas expulsion (i.e. $\sfe\geq0.15$), and we consider the efficiencies $\sfe=0.15$, 0.17, 0.20, 0.25,  and birth masses $M_\star=3$k, 6k, 10k, 15k, 30k, 60k $\msun$ (equivalent to a number of stars $N_\star\approx[5\times10^3,\times10^5]$). We do not consider any eccentric or inclined (with respect to the Galactic disk plane) orbits in this study.

\begin{figure}
\centering
 \includegraphics[width=0.5\textwidth]{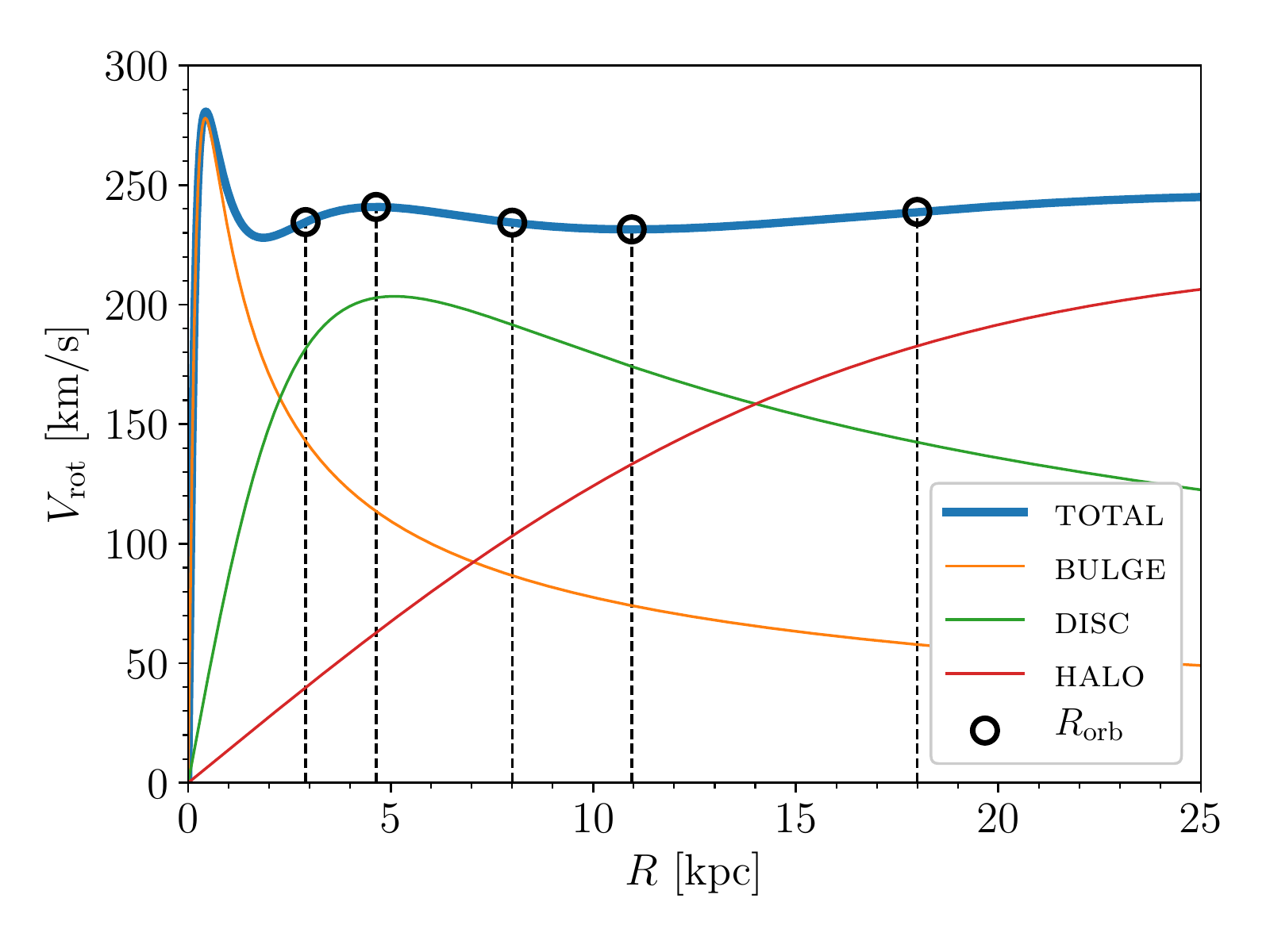} 
 \caption{The rotation curve of our Galaxy model (thick blue line) and its corresponding components (bulge, disc and halo). The black open circles show the radii of the circular orbits on which we put our model clusters: $R_\mathrm{orb}=2.9$, 4.64, 8.0, 10.95, 18.0~kpc.}
 \label{fig:vrot}
\end{figure}

In the scope of this paper we vary the impact of the tidal field, characterized by the ratio of the cluster half-mass radius to Jacobi (tidal) radius, $\rhrt$, in two ways with respect to our standard set of models: 
\begin{itemize}
  \item[1)] We vary the Galactocentric distance $R_{\tx[orb]}$ of the model clusters while keeping constant their physical size ($r_{\tx[h]}$) at the time of instantaneous gas expulsion. We have chosen 4 additional Galactocentric distances: $R_\mathrm{orb}~=~2.9$, 4.64, 10.95, and 18.0~kpc, which correspond to $\rhrt=0.1,$ 0.075, 0.04, and 0.03. This completes our initial set of ``standard'' S0-model clusters, i.e. $\rhrt=0.052$ at $R_\mathrm{orb}=8.0\ \tx[kpc]$. 
  We name the additional models as `extra Inner-' (xI), `Inner-' (I), `Outer-' (O) and `extra Outer-' (xO) orbit model clusters, respectively.
  In Fig. \ref{fig:vrot}, which presents the rotation curve of our Galaxy model, their positions are indicated by black open circles. 

  \item[2)] We vary the physical size $r_{\tx[h]}$ of the 'S0-model' clusters, while keeping them in the solar neighbourhood ($R_{\tx[orb]}=8.0\ \tx[kpc]$). 
  That is, we vary the cluster density.
   In this case, we expand the few simulations already performed in \PaperI\ for $\rhrt=0.1$, 0.075, 0.025 {into} a larger range of birth masses and more realizations per model. These models are named: `the most diffuse' (or S+2, when $\rhrt=0.1$), `the diffuse' (or S+1, 0.075) and `the compact' (or S$-1$, 0.025) model clusters.
\end{itemize}

The number of realizations performed per model, where each model is described by a global SFE ($\sfe$), cluster birth mass $M_\star$ and impact of the tidal field $\rhrt$, are presented in Table \ref{tab:models}.

\begin{table*}
  \centering
  \caption{Number of realizations performed for each model cluster, where each model is described by its birth mass ($M_\star$), global SFE ($\sfe$) and impact of the tidal field ($\rhrt$).}
  \label{tab:models}
  \begin{tabular}{rrrrrrrrrr}
  \hline
  $M_\star/\msun$&$\sfe$&&&&$n_{\tx[rnd]}$&&&& \\
  \hline 
 &  $\rhrt = $ &0.100 (S+2)  &0.100 (xI)  &0.075 (I)  &0.070 (S+1)  &0.050 (S0)  &0.040 (O)  &0.030 (xO)  &0.025 (S-1)\\
 \hline\hline
  3000 & 0.15 & 15 & 15 & 16 & 15 & 26 & 16 & 15 & 15\\
  3000 & 0.17 &  6 &  6 &  6 &  6 &  6 &  6 &  6 &  6\\
  3000 & 0.20 &  6 &  6 &  6 &  6 &  6 &  6 &  6 &  6\\
  3000 & 0.25 &  1 &  1 &  1 &  1 &  1 &  1 &  1 &  1\\
  6000 & 0.15 & 10 & 10 & 11 & 10 & 36 & 11 & 10 & 10\\
  6000 & 0.17 &  6 &  6 &  6 &  6 &  6 &  6 &  6 &  6\\
  6000 & 0.20 &  8 &  8 &  6 &  8 &  8 &  6 &  8 &  8\\
  6000 & 0.25 &  3 &  3 &  1 &  3 &  3 &  1 &  3 &  3\\
 10000 & 0.15 & 11 & 11 & 11 & 11 & 36 & 11 & 11 & 11\\
 10000 & 0.20 &    &    &    &    &  1 &    &    &   \\
 10000 & 0.25 &    &    &    &    &  1 &    &    &   \\
 15000 & 0.15 & 11 & 11 & 12 & 11 & 32 & 12 & 11 & 11\\
 15000 & 0.17 &  6 &  6 &  6 &  6 &  6 &  6 &  6 &  6\\
 15000 & 0.20 &  8 &  8 &  6 &  8 &  8 &  6 &  8 &  8\\
 15000 & 0.25 &  3 &  3 &  1 &  3 &  3 &  1 &  3 &  3\\
 30000 & 0.15 & 10 & 10 & 11 & 10 & 27 & 11 & 10 & 10\\
 30000 & 0.17 &  6 &  6 &  6 &  6 &  6 &  6 &  6 &  6\\
 30000 & 0.20 &  6 &  6 &  6 &  6 &  6 &  6 &  6 &  6\\
 30000 & 0.25 &  1 &  1 &  1 &  1 &  1 &  1 &  1 &  1\\
 60000 & 0.15 &    &  8 & 15 &    & 15 & 15 &    &   \\
 60000 & 0.17 &    &    &  1 &    &  1 &  1 &    &   \\
 60000 & 0.20 &    &    &  1 &    &  1 &  1 &    &   \\
 60000 & 0.25 &    &    &  1 &    &  1 &  1 &    &   \\
100000 & 0.15 &    &    &    &    &  3 &    &    &   \\
300000 & 0.15 &    &    &    &    &  1 &    &    &   \\
\hline
\end{tabular}

\end{table*}

\section{The violent relaxation duration}\label{sec:VR}

The impact of the tidal field on the cluster survivability can be quantified by the variations of the cluster bound mass fraction,$M_\mathrm{J}/M_\star$, at the end of violent relaxation, (i.e. $t=\tvr$), or final bound fraction, 
\begin{equation}
F_\mathrm{bound} = \frac{M_\mathrm{J}(t=\tvr)}{M_\star}.
\end{equation}

{Originally, in \PaperI\ we assumed that the violent relaxation ends at an age of 20 Myr\footnote{i.e. 20 Myr after the instantaneous gas expulsion}, a limit which we defined by visual inspection as the time when the rapid decrease of the bound mass fraction of all model clusters stops and turns into a constant (see Fig. 3 of \PaperI). We then reported that the duration of the violent relaxation of solar neighbourhood clusters does not differ significantly for different impacts of the tidal field. In \citet{Bek+18}, when we studied the long-term evolution of our model clusters, we shifted our definition of the end of violent relaxation to 30~Myr, to be sure it is totally over.
But now, that we consider the impact of the tidal field also at different Galactocentric distances, we need a more robust definition of the violent relaxation duration, $\tvr$. 
We define it as the time when the total mass-loss rate of a cluster becomes equal to its stellar evolutionary mass-loss rate. Since the total mass-loss rate of a violently relaxing cluster consists of both stellar evolutionary mass-loss and the escape of unbound stars due to violent relaxation our criterion for $\tvr$ implies that violent relaxation is actually over.}


{In order to measure the mass-loss rates of star clusters we introduce their time-scale for total mass-loss, defined as the inverse of the cluster total mass-loss rate normalized to the cluster birth mass
\begin{equation}
\tau = \left({-\displaystyle \frac{\tx[d]M_{\tx[J]}/M_\star}{\tx[d]t}}\right)^{-1} = -\frac{M_\star}{\dot{M_{\tx[J]}}}.
\label{eq:tau}
\end{equation}
Similarly, the time-scale for stellar evolutionary mass loss is defined as the inverse of the cluster stellar evolutionary mass-loss rate normalized to the cluster birth mass 
\begin{equation}
\tau_\mathrm{stev} = -\displaystyle\frac{M_\star}{\dot{M}_\mathrm{stev}}.
\label{eq:taustev}
\end{equation}
}

\citet{Lamers+05a} already provided some approximation, which describes the stellar evolutionary mass loss fraction of their model clusters for ages $t>12.5$~Myr with an accuracy of a few per cent. 
However, we need an expression of the stellar evolutionary mass-loss which is valid also at younger ages.
Additionally, we use the IMF of \citet{K2001}, instead of the IMF of \citet{Salpeter55} as \citet{Lamers+05a} did. 
{Therefore, we fit our own approximation for $\tau_\mathrm{stev}$, which provides an excellent fit for the age range from 4 to 100~Myr}\footnote{{$\tau_\mathrm{STEV}$ is the same for any cluster mass, except for a large background noise in the case of low-mass clusters.}}
\begin{equation}
\tau_\mathrm{stev} = 17.8\left({t-2}\right)^{1.06}\quad{[\mathrm{Myr}]}.
\label{eq:taufit}
\end{equation}

\begin{figure*}
\centering
\begin{tabular}{cc}
(a) $R_\mathrm{orb} = 2.9$ kpc, $t_\mathrm{VR} = 7.9\pm1.1$ Myr & (b) $R_\mathrm{orb} = 18.0$ kpc, $t_\mathrm{VR} = 33.0\pm4.6$ Myr \\
\includegraphics[width=0.5\textwidth]{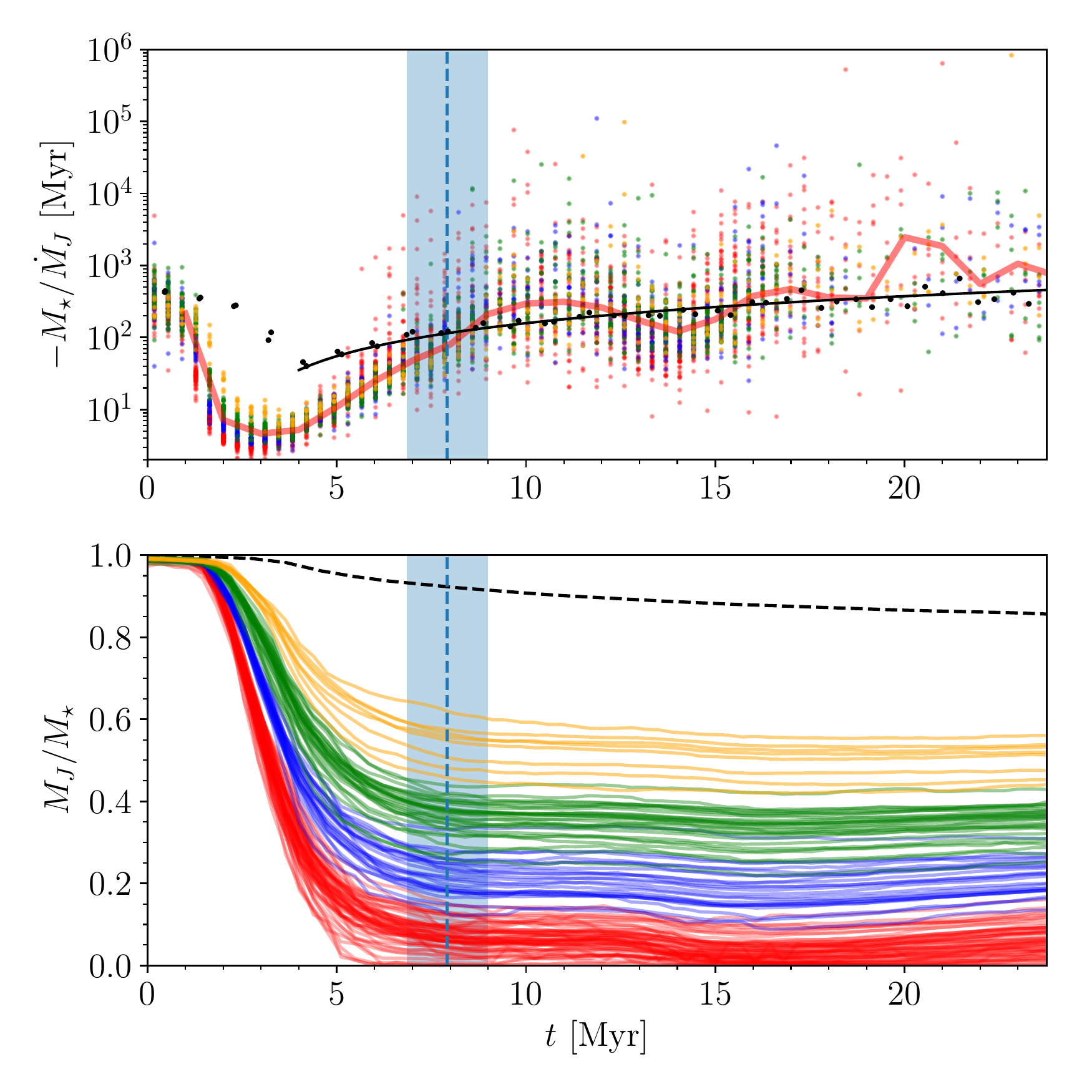} &  \includegraphics[width=0.5\textwidth]{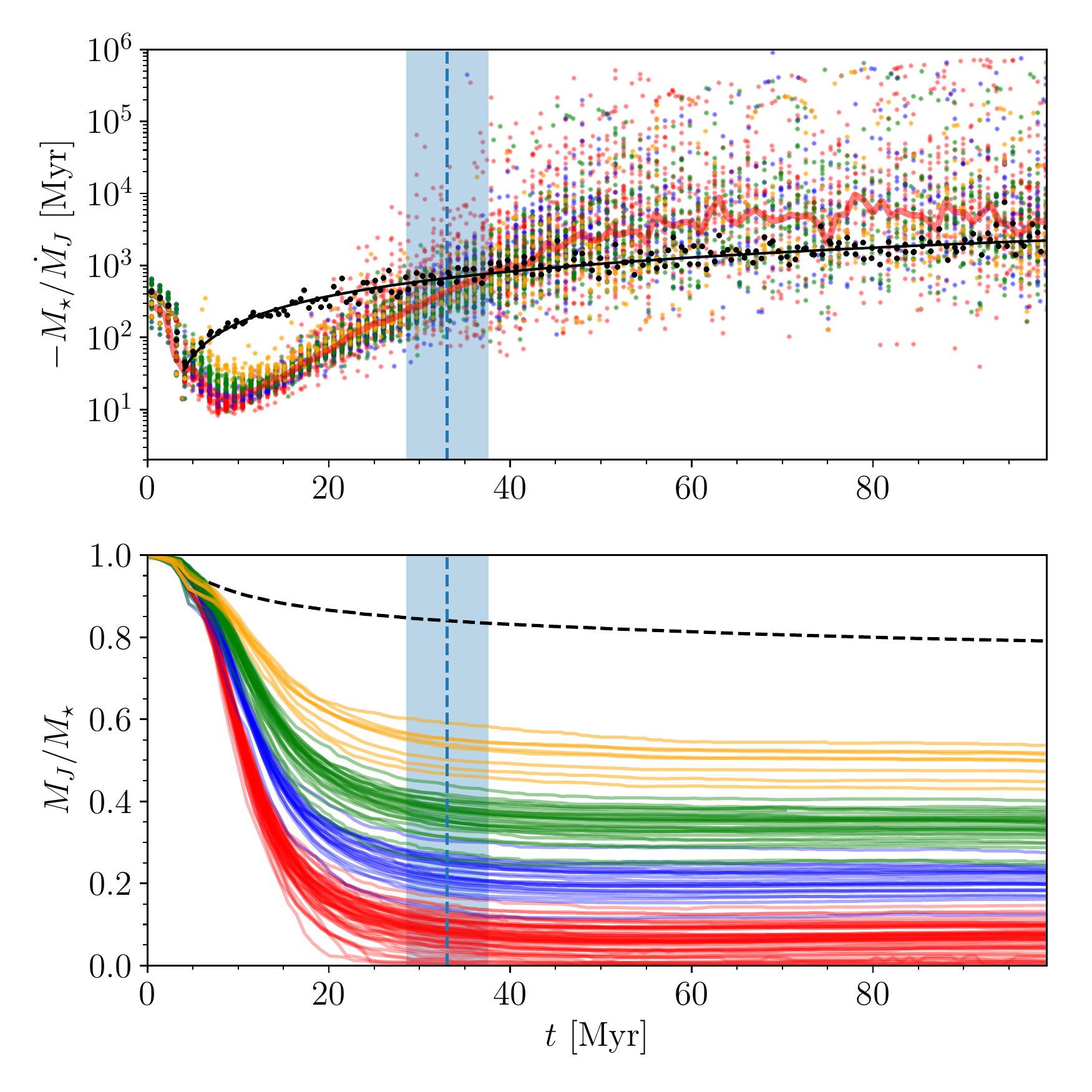} \\
\end{tabular}
 \caption{The mass-loss time-scale (top panels) and the bound mass fraction evolution (bottom panels) of star clusters at two Galactocentric distances $R_\mathrm{orb} = 2.9$~kpc and $R_\mathrm{orb}=18$~kpc are shown in left and right panels, respectively. 
 The different colours code the cluster global SFEs: $\sfe = 0.15$ (red), 0.17 (blue), 0.20 (green), and 0.25 (yellow).
 The red thick line in the upper panels corresponds to the median of all simulations at a given time (irrespective of global SFE and birth mass).
 The black dots show the time-scale for stellar evolutionary mass-loss and correspond to two simulations with $M_{\star}=10^5\msun$. The black solid line is the corresponding best fit.
 In the bottom panels the black dashed line shows the stellar evolutionary mass-loss of a $M_{\star}=10^5\msun$ cluster.
 In each panel the vertical blue dashed line corresponds to the end of violent relaxation, $t_\mathrm{VR}$, with the shaded area corresponding to the standard deviation (see the text for more explanations).}
 \label{fig:masslosstime}
\end{figure*}

In top panels of Fig. \ref{fig:masslosstime} we present the evolution with time of the time-scale for total mass-loss, $\tau$, of our model clusters.
We show here as examples clusters at two Galactocentric distances, $R_{\tx[orb]}=2.9$~kpc and $R_{\tx[orb]}=18.0$~kpc in the left and right panels of Fig. \ref{fig:masslosstime}, respectively. 
In the bottom panels of Fig. \ref{fig:masslosstime} the corresponding bound mass fraction evolutions are presented. The different colours correspond to different global SFEs (red, blue, green and yellow for $\sfe=0.15,$ 0.17, 0.20 and 0.25, respectively). 
The red thick curve in the top panels shows the median value of the total mass-loss time-scales of all model clusters for a given environment (i.e. $R_\mathrm{orb}$). 
The black dashed line in the bottom panels depicts stellar evolutionary mass-loss. 
The stellar evolutionary mass-loss time-scales as defined by Eq. \ref{eq:taustev} for two $M_\star=10^5 \msun$ cluster models are shown with black dots in top panels, with the thin solid black curve being the corresponding best fit (Eq. \ref{eq:taufit}). 

We compare the total mass-loss time-scale, $\tau$, of all models with the fit function for stellar evolution provided in Eq. \ref{eq:taufit}. That is, we identify the time when the total mass-loss time-scale of our model clusters becomes longer than the fit function. We then take the average over all model clusters for a given impact of the tidal field and define it as the end of violent relaxation. This average $\tvr$ is indicated by the vertical dashed line in each panel of Fig \ref{fig:masslosstime}, where the shaded area corresponds to the standard deviation. {We provide our measurements of the end of violent relaxation, $\tvr$, as a function of the impact of the tidal field, $\lambda$, in Table~\ref{tab:VRend}.}

{
Table~\ref{tab:VRend} consists of two parts: the first-half corresponds to different Galactocentric distances, while the second-half corresponds to the solar neighbourhood clusters with different sizes. As we can see from Table~\ref{tab:VRend} our new, more robust measurements of the end of violent relaxation for the local clusters are consistent with the previous estimates in \PaperI. }

\begin{table}
  \centering
  \caption{End of violent relaxation for different impacts of the tidal field.}
  \label{tab:VRend}
  \begin{tabular}{cccc}
  \hline
  Model name &  $R_\mathrm{orb}$ [kpc] &$\rhrt$ & $t_\mathrm{VR}$ [Myr]\\
  \hline\hline
xI  &  2.9  & 0.100  & $  7.9 \pm 1.1 $ \\
I   &  4.64 & 0.075  & $ 12.4 \pm 2.3 $ \\
O   & 10.95 & 0.040  & $ 23.9 \pm 2.9 $ \\
xO  & 18.0  & 0.030  & $ 33.0 \pm 4.6 $ \\
\hline
S+2 &  8.0  & 0.100  & $ 19.5 \pm 3.2 $ \\
S+1 &  8.0  & 0.070  & $ 17.5 \pm 1.9 $ \\
S0  &  8.0  & 0.050  & $ 17.9 \pm 2.3 $ \\
S-1 &  8.0  & 0.025  & $ 14.4 \pm 1.7 $ \\
\hline
\end{tabular}

\end{table}

{
The duration of violent relaxation, $\tvr$, correlates linearly with the Jacobi radius, $R_\mathrm{J}$, when we consider clusters with the same size and mass at different Galactocentric distances. 
This correlation illustrates the longer time-span needed by escaping stars to reach the Jacobi radius when considering clusters with identical velocity dispersion but larger Galactocentric distances (first half of Table~\ref{tab:VRend}).
}

{When we consider local clusters, compact clusters have shorter violent relaxation than diffuse ones (second part of Table~\ref{tab:VRend}). This is because, more compact clusters have a higher velocity dispersion, allowing their escaping stars to reach the Jacobi radius faster than those of diffuse clusters.}

\section{Final bound fraction}\label{sec:FB}

The final bound fraction $\fb$ is the cluster birth mass fraction remaining gravitationally bound to the cluster at the end of violent relaxation, i.e. at $t=\tvr$. 
This is an indicator of cluster survivability after gas expulsion.
We showed previously that the final bound fraction does not depend on the cluster birth mass $M_\star$ for solar neighbourhood clusters (\PaperI, \PaperII). This stands also for other galactocentric distances \citep{BekThesis}.
Therefore,  in the following figures, the final bound fraction for a given global SFE and a given impact of the tidal field has been averaged through all cluster birth masses, and their corresponding realizations.

\begin{figure}
\centering
  \includegraphics[width=0.5\textwidth]{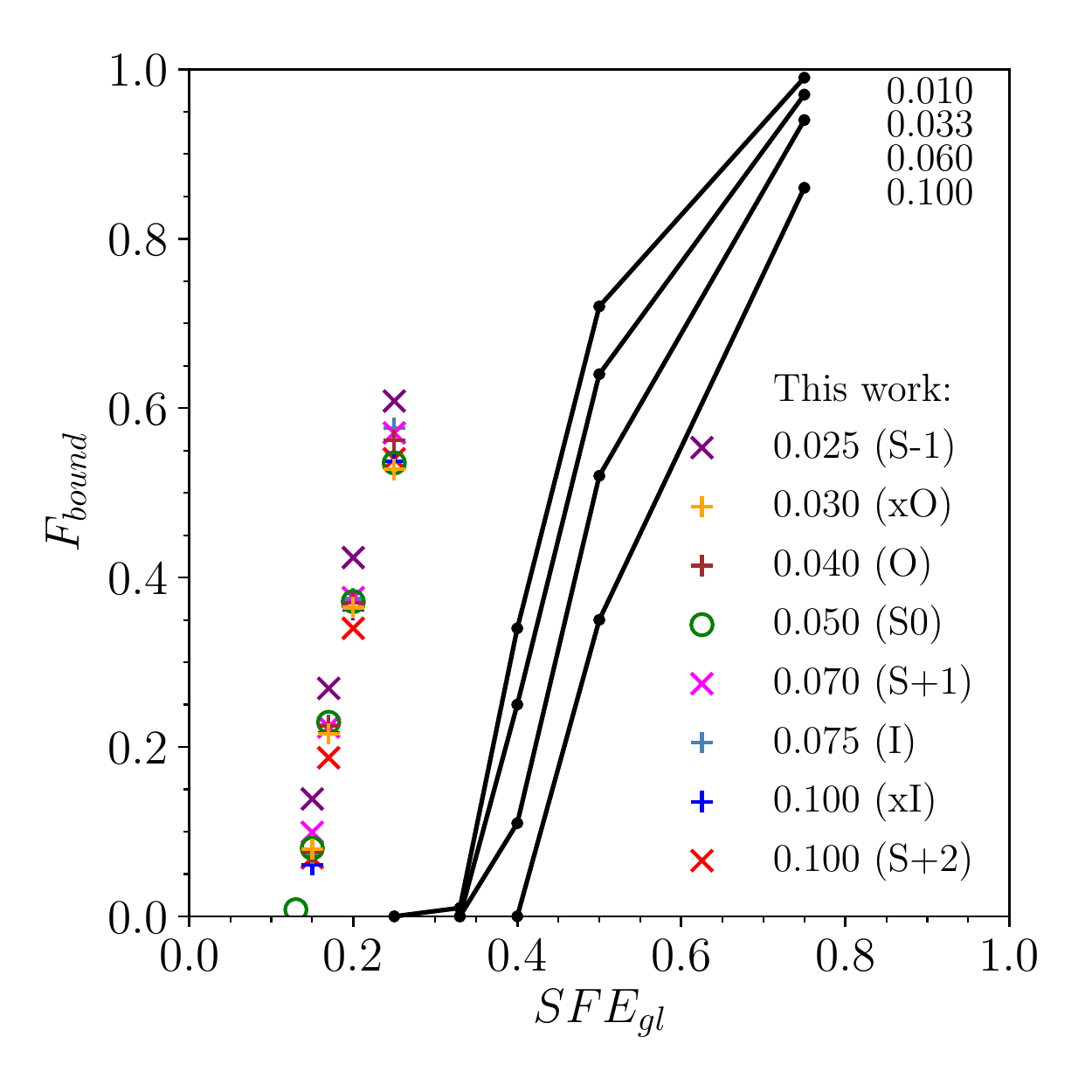} 
  \caption{The final bound fraction as a function of global SFE and impact of the tidal field. The different colours correspond to different impacts of the tidal field. Green open circles show our S0-models. The solar neighbourhood clusters with different half-mass radii are shown by cross symbols, while clusters at different Galactocentric distances are presented by plus symbols. Each point corresponds to the mean bound fraction of a set of simulations with a given global SFE and $\rhrt$. Each set of simulations consists of models with birth masses ranging from 3k to 30k $\msun$ at least. The black lines correspond to the results of \citet{BK07} for the case of instantaneous gas expulsion with their impacts of the tidal field ($r_\mathrm{h}/t_\mathrm{t}$) shown next to the black dots. }
  \label{fig:fb-sfe}
 \end{figure}

Figure \ref{fig:fb-sfe} presents the final bound fraction as a function of the global SFE and of the impact of the tidal field. 
Standard deviations are not shown for the sake of clarity. 
Our standard S0-model clusters are indicated by green open circles. Solar neighbourhood clusters of different sizes are represented by `$\times$'-symbols, while model clusters at different Galactocentric distances are indicated by `$+$'-symbols.
In each case, the colour-coding corresponds to different impacts of the tidal field (see the key).
We compare our results with those of \citet{BK07} obtained for instantaneous gas expulsion and different impacts of the tidal field $r_\mathrm{h}/r_\mathrm{t} = 0.01,$ 0.033, 0.06 and 0.100 (black dots connected with lines, from top to bottom). 

As we mentioned before, \citet{BK07} considered clusters with a radially constant SFE, on circular orbits in a spherical gravitational potential representative of the host galaxy. 
In contrast, we consider clusters formed with a centrally-peaked SFE profile, as a result of star-formation taking place with a constant efficiency per free-fall time, and moving on circular orbits in the disk plane of an axisymmetric Galactic potential consisting of a bulge,  a disk and a dark halo. 
{In both cases, the stellar component of a gas embedded cluster has a Plummer density profile and is in virial equilibrium immediately before gas expulsion. 
The main difference between both models, if they were considered to be isolated, would be in the virial ratio (the ratio between kinetic and potential energies) profile of their stellar component. That is, in the case of \citet{BK07}, the virial ratio of stars immediately after gas expulsion is about constant through the cluster, because the fraction of stars and gas was constant. In our model clusters, however, the virial ratio profile decreases toward the cluster centre, because of the centrally-peaked SFE profile. 
Due to this difference, as it is shown in Fig. \ref{fig:fb-sfe}, not only are our model clusters able to resist instantaneous gas expulsion on SFE as low as $\sfe=0.15$, their survival likelihood depends weakly only on the impact of the tidal field.}

{In other words, the centrally-peaked SFE profile, by dampening the cluster expansion, helps prevent its destruction by the tidal field of the Galaxy, even for large values of $\lambda$ (e.g. $\lambda=0.1$).
Decreasing the impact of the tidal field will not help save a greater bound fraction. It will only delay the end of violent relaxation, giving more time for unbound stars to leave the cluster \citep[see Table \ref{tab:VRend} and also Fig. 27 of][]{BekThesis}. }

\begin{figure}
\centering
  \includegraphics[width=0.5\textwidth]{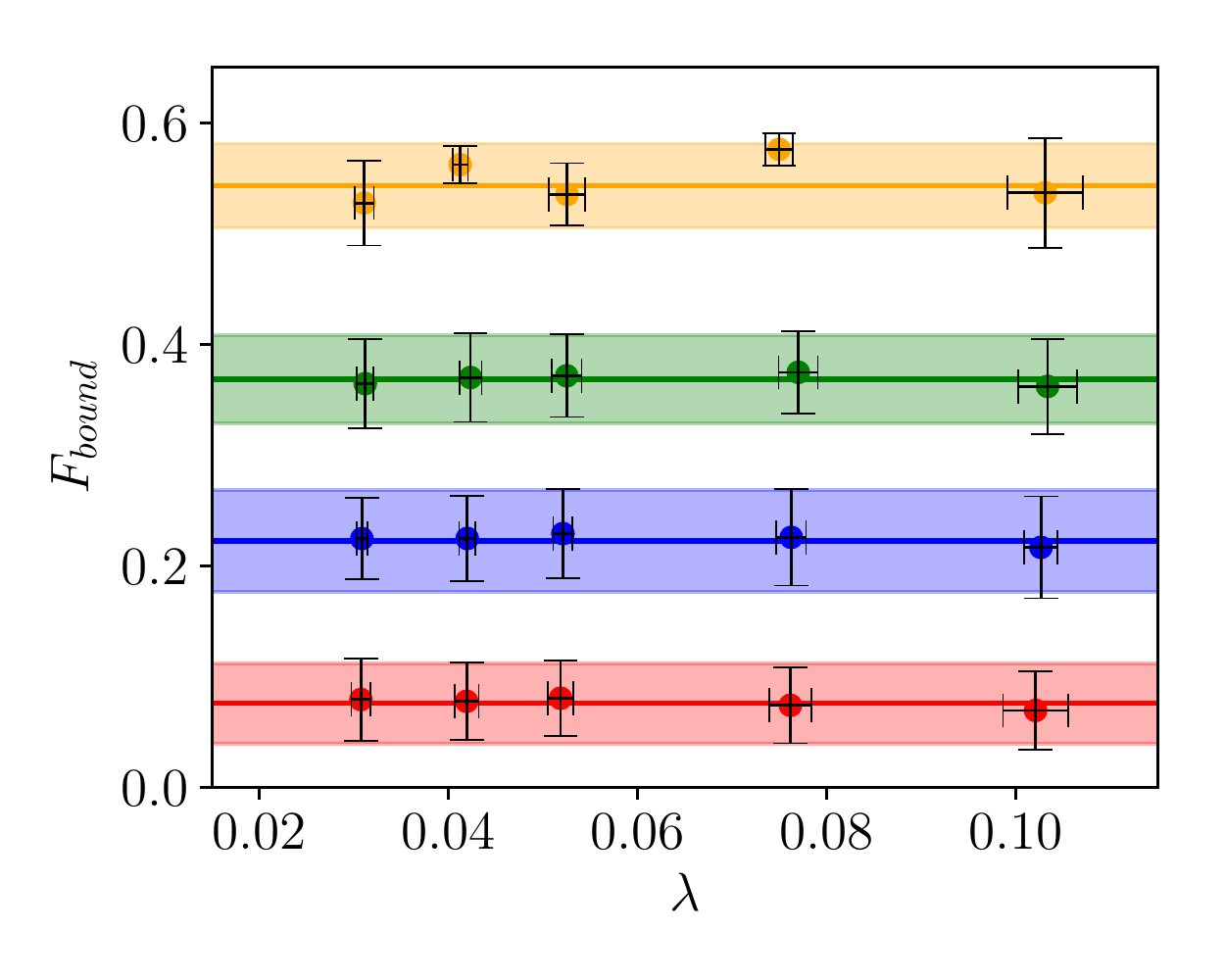} 
  \caption{The final bound mass fraction as a function of the impact of the tidal field for different Galactocentric distances. The colour-coding corresponds to the global SFE and is the same as in Fig. \ref{fig:masslosstime}. Each point corresponds to the mean and standard deviation of model clusters with the same global SFE and impact of the tidal field.}
  \label{fig:envr}
\end{figure}

Figure \ref{fig:envr} presents the bound fraction of our model clusters against the impact of the tidal field for different galactocentric distances. 
{Different colours correspond to different global SFEs,($\sfe=$0.25, 0.20, 0.17 and 0.15, from top to bottom). Coloured lines with shaded areas correspond to the mean final bound fraction averaged over all $M_\star$ and $\lambda$ for a given $\sfe$, and the corresponding standard deviation. 
This figure} demonstrates, that the final bound fraction is not affected by the tidal field when we consider different Galactocentric distances. 

\begin{figure}
\centering
  \includegraphics[width=0.5\textwidth]{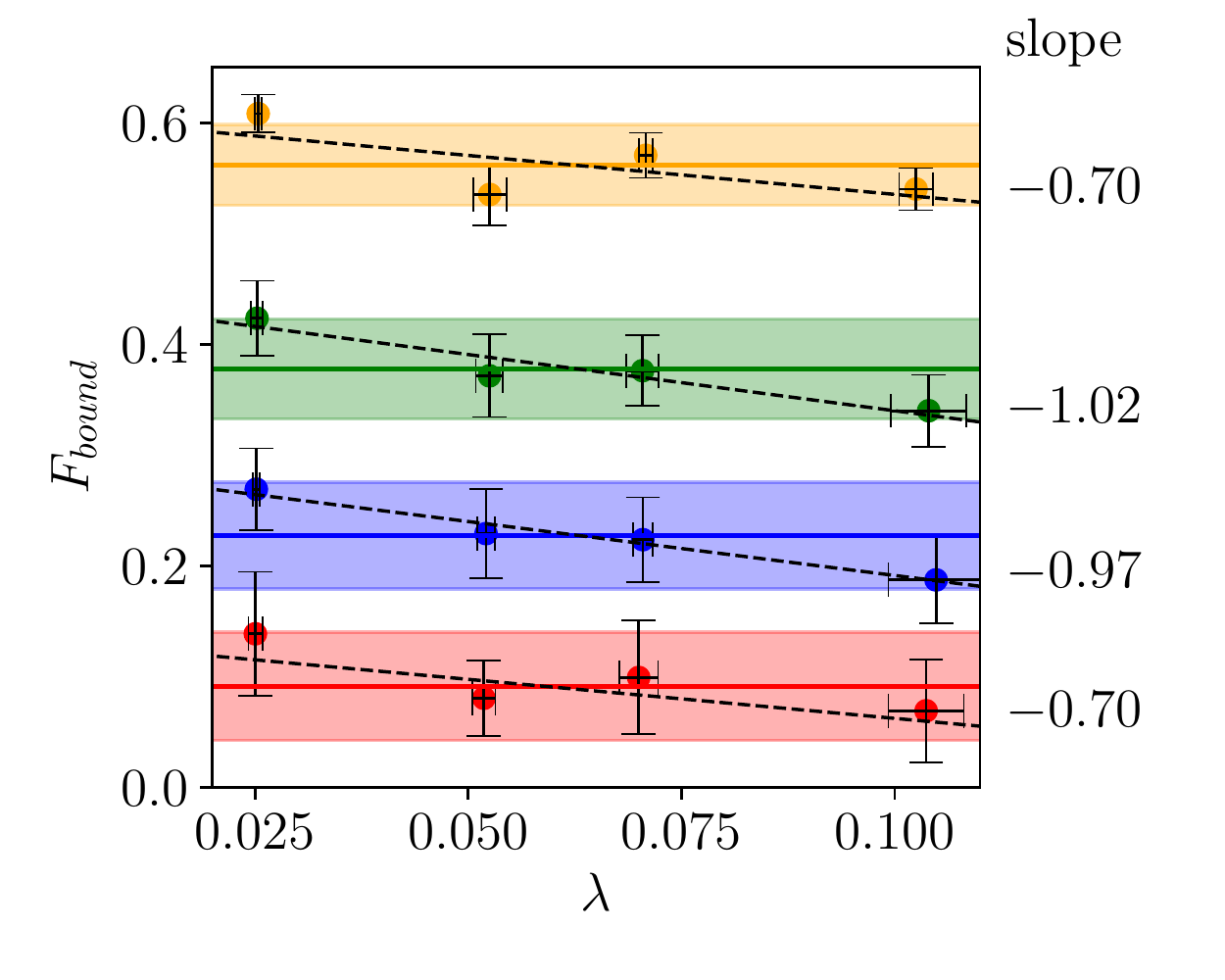} 
  \caption{The bound mass fraction at the end of violent relaxation as a function of the impact of the tidal field for different cluster central densities at $R_{\tx[orb]}~=~8.0~\tx[kpc]$. Different colours correspond to different global SFEs as in Figs. \ref{fig:masslosstime} and \ref{fig:envr}. Each point corresponds to the mean and standard deviation for clusters with the same global SFE and impact of the tidal field. }
  \label{fig:Senvr}
 \end{figure}

Figure~\ref{fig:Senvr} presents the final bound fraction as a function of the impact of the tidal field, but for a fixed Galactocentric distance of $R_\mathrm{orb}=8\tx[\ kpc]$. {In this case, a decreasing trend can be seen, although the differences in the final bound fraction remain consistent with each other within the error-bars.} The black dashed lines are the linear fits to the final bound fractions of a given global SFE as a function of $\rhrt$. Their slopes are shown on the right-hand-side of the figure {and demonstrate that the trend is weak. A larger range of the impact of the tidal field is needed to ascertain it.} 

We therefore conclude that the survivability of our model star clusters after instantaneous gas expulsion is independent of the impact of the tidal field, regardless of whether this one is due to a varying cluster size or a varying cluster Galactocentric distance.

\section{Conclusions}\label{sec:conc}

We have studied the impact of the tidal field on the survivability of star clusters after instantaneous gas expulsion. To do so, we have expanded our grid of simulations from \PaperI\ and \PaperII, and considered different Galactocentric distances ($R_\mathrm{orb}=2.9,$ 4.64, 8.0, 10.95, 18.0~kpc), as well as solar neighbourhood clusters (i.e. $R_\mathrm{orb}=8\ \mathrm{kpc}$) with different volume densities. Both cases yield variations of the impact of the tidal field, that we define as $\rhrt = r_\mathrm{h}/R_\mathrm{J}$. 

Our model star clusters are formed with a centrally-peaked SFE profile and have circular orbits in the Galactic disc plane, with the Galactic potential modelled as a three-component axisymmetric Plummer-Kuzmin model \citep{MiyamotoNagai75,Just+09}.

We have measured the duration of violent relaxation for all our model clusters. We define the end of violent relaxation as the moment when stellar evolutionary mass losses start dominating the rapid (violent) mass-loss resulting from gas expulsion. 
As we showed previously in \PaperI\ the violent relaxation duration does not depend significantly on the cluster global SFE and birth mass. 
In our simulations the violent relaxation does not last longer than 50 Myr. {It is shorter for clusters closer to the Galactic centre and nearly constant when $R_\mathrm{orb}=8.0$~kpc is fixed, slightly decreasing with $\lambda$ (see Table \ref{tab:VRend}).}

Next, we have measured the final bound fraction (i.e. bound mass fraction at the end of violent relaxation) of our clusters to quantify the impact of the tidal field on the cluster survivability.
Identical clusters located at different Galactocentric distances, which thus experience different impacts of the tidal field, show the same final bound fraction at the end of violent relaxation irrespective of their galactocentric distance (Fig. \ref{fig:envr}). However, clusters at the same Galactocentric distance of 8 kpc, but with different volume densities, and therefore different impacts of the tidal field as well, present small variations of about 0.1 in their final bound fraction such that compact clusters retain a slightly higher fraction of their stars than diffuse clusters.

In general, we conclude that, within the scope of our simulations, the cluster survivability after instantaneous gas expulsion as measured by their bound mass fraction at the end of violent relaxation $F_\mathrm{bound}$, is independent of the impact of the tidal field $\rhrt$. 

\section*{Acknowledgements}
This work was supported by Sonderforschungsbereich SFB 881 ``The Milky Way System'' (subproject B2) of the German Research Foundation (DFG). 
The authors gratefully acknowledge Prof.~Walter~Dehnen (University of Leicester, UK) for his support and discussions in connection with implementing the code \textsc{mkhalo} for our purposes. 
B.S. and G.P. gratefully acknowledge Prof.~Rupali~Chandar (University of Toledo, OH, USA) for stimulating discussions.  
B.S. gratefully acknowledges Prof.~Rainer~Spurzem for his support with accessing the high-performance computing clusters at JURECA and LAOHU.
B.S. acknowledges the support within PCF program BR05236322 funded by the Ministry of Education and Science of the Republic of Kazakhstan.
P.B. and B.S. acknowledge the support of the Volkswagen Foundation under the Trilateral Partnerships grant 90411 and the support by the National Astronomical Observatories of Chinese Academy of Science (NAOC/CAS) through the Silk Road Project, through the Thousand Talents (``Qianren'') program and (P.B. only) the President's International Fellowship for Visiting Scientists and the National Science Foundation of China under grant No. 11673032.
P.B. acknowledges the special support by the NASU under the Main Astronomical Observatory GRID/GPU computing cluster project. This work benefited from support by the International Space Science Institute, Bern, Switzerland, through its International Team program ref. no. 393 ``The Evolution of Rich Stellar Populations \& BH Binaries'' (2017--18).
We acknowledge the use of supercomputers of the J\"ulich Supercomputing Centre (JSC) JURECA (hhd28), of the Baden-W\"urttemberg HPC infrastructure (bwforCluster, national and state funded through grant INST 35/1134-1 FUGG, and funded by SFB 881), of the KEPLER GPU cluster funded by Volkswagen foundation project I/81 396, and of NAOC/CAS (LAOHU at Centre of Information and Computing).




\bibliographystyle{mnras}
\bibliography{paper3} 








\bsp	
\label{lastpage}
\end{document}